\documentclass[letterpaper]{jpconf}

\usepackage{epsfig}
\usepackage{iopams} 
\usepackage{amstext,amsbsy,amsgen,amscd,amsopn,amsfonts,amssymb,amsbsy}
\usepackage{nicefrac}
\usepackage{hyperref}

\hypersetup{pdftitle={SUSY on the Rocks},
pdfsubject={Proceedings of TeV Particle Astrophysics II},
pdfauthor={Markus Tobias Ahlers},
plainpages=false,
pdfpagemode=None,
pdfpagelayout=OneColumn,
pdfcenterwindow=true,
pdfview=FitV,
pdfstartview=FitV,
pdffitwindow=true,
colorlinks=true,
bookmarksopen=false,
bookmarksnumbered=true,
bookmarksopenlevel=2,
pdfmenubar=true,
pdfwindowui=true,
pdffitwindow=true}

\newcommand{\D}{{\rm d}}
\begin{document}
\title{Supersymmetry on the Rocks{\hfill\normalsize\tt DESY 06-193}}

\author{Markus Ahlers}

\address{Deutsches Elektronen-Synchrotron DESY, Notkestra\ss e  85, 22607 Hamburg, Germany}

\ead{markus.ahlers@desy.de}

\begin{abstract}
In $\mathcal{R}$-parity conserving supersymmetric (SUSY) models the lightest SUSY particle (LSP) is stable and a candidate for dark matter. Depending on the coupling and mass of this particle the life time of the next-to-lightest SUSY particle (NLSP) may be large compared to experimental time scales. In particular, if the NLSP is a charged particle and its decay length is of the order of the Earth's diameter Cherenkov telescopes might observe parallel muon-like tracks of NLSP pairs produced in neutrino-nucleon interactions in the Earth's interior. We have investigated two SUSY scenarios with a long-lived $\widetilde{\tau}$ NLSP and a gravitino LSP in view of the observability at the IceCube detector.
\end{abstract}

\section{Introduction}

The interactions of high energy ($E\gtrsim 1$~PeV) cosmic particles with the Earth matter are a probe of physics beyond the TeV scale. It was pointed out by the authors of Ref.~\cite{Albuquerque:2003mi} that under favorable conditions a hypothetic charged particle emerging from neutrino-nucleon interactions in the Earth's interior could give rise to an appreciable rate of muon-like tracks in Cherenkov telescopes. The underlying flux of high energy neutrinos is expected from photopion processes of high energy protons with the cosmic microwave background (cosmogenic neutrinos)~\cite{Beresinsky:1969qj,Stecker:1978ah} or with the ambient photon gas of cosmic accelerators~\cite{Waxman:1998yy,Ahlers:2005sn}. For the observability of these new particles at neutrino telescopes the low production cross section and the small branching ratio compared to Standard Model (SM) processes has to be balanced by a sufficiently long life time and a low energy loss in matter.

\section{Supersymmetry with a Long-lived Stau NLSP}

Attractive candidate scenarios are supersymmetric (SUSY) extensions of the SM with $\mathcal{R}$-parity conservation. In these models the next-to-lightest super-particle (NLSP) can only decay into final states containing the lightest super-particle (LSP) which is stable. For a gravitino LSP with mass $m_{\nicefrac[\textrm]{3}{2}}$ the decay length $L$ of the NLSP in units of the Earth's diameter $2R_\oplus$ is approximately~\cite{Giudice:1998bp}
\begin{equation}
\bigg(\frac{L}{2 R_\oplus}\bigg) \approx \bigg(\frac{m_{\text{NLSP}}}{100\, \text{GeV}}\bigg)^{-6}\bigg(\frac{m_{\nicefrac[\textrm]{3}{2}}}{400\, \text{keV}}\bigg)^2\bigg(\frac{E_{\text{NLSP}}}{500\, \text{GeV}}\bigg).
\end{equation}
Constraints from big bang
nucleosynthesis and the cosmic microwave background yield an upper limit on the mass of a gravitino LSP between $10$ and $100$~GeV~\cite{Cerdeno:2005eu}. Throughout our calculations we will assume that the NLSP is the right-handed stau ($\widetilde{\tau}_\text{R}$), the supersymmetric scalar partner of the right-handed tau, and treat this particle as effectively stable corresponding to a gravitino mass $m_{\,\nicefrac[\textrm]{3}{2}}\gtrsim400$~keV. 
 
Due to $\mathcal{R}$-parity conservation neutrino-nucleon interactions will allways produce pairs of super-particles, which then promptly decay into long-lived $\widetilde{\tau}_\text{R}$ NLSPs. It was argued in Ref.~\cite{Ahlers:2006pf} that the rate of single stau events is subdominant compared to the flux of muons, even for the most optimistic case of very light squarks with masses around $300$~GeV. This is connected to the fact that tracks of high energy staus in a Cherenkov detector will look like low energy muons which have a much larger rate. Instead, a promising signal could be the observation of stau pairs which might show up as parallel tracks in the detector due to their large boost~\cite{Albuquerque:2003mi}. In the following sections we will give a short review of the calculation of this $\widetilde{\tau}\!+\!\widetilde{\tau}$ flux through a neutrino telescope such as IceCube~\cite{Ahrens:2002dv} and comment on some phenomenological aspects.

\begin{figure}[t!]
\begin{center}
\begin{minipage}[t]{0.47\linewidth}
\includegraphics[width=\linewidth,clip=true]{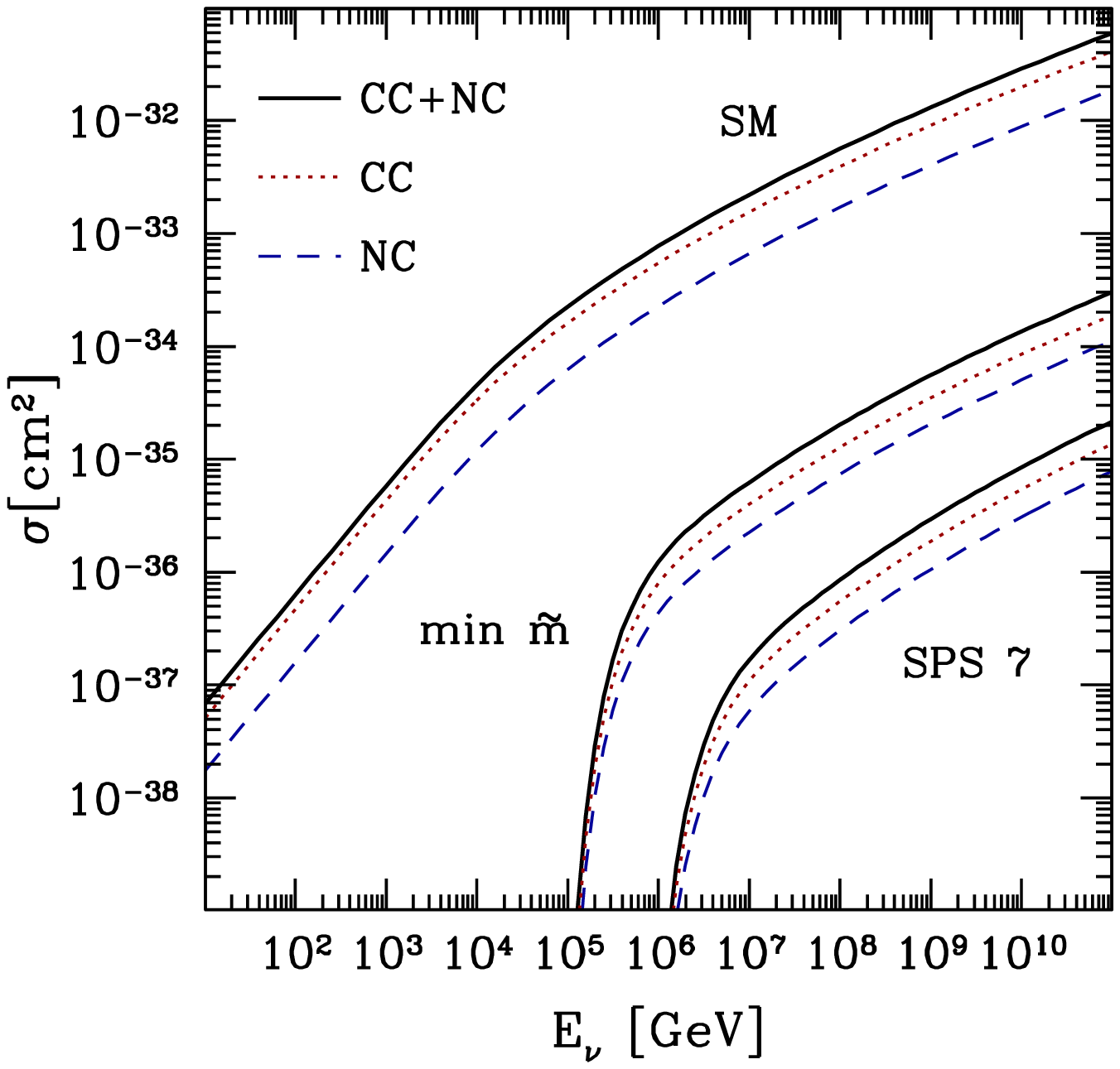}
\caption[]{\label{sigma}The neutrino nucleon cross section of the SM compared to contributions from neutralino and chargino exchange.}
\end{minipage}
\hfill
\begin{minipage}[t]{0.47\linewidth}
\includegraphics[width=\linewidth,clip=true]{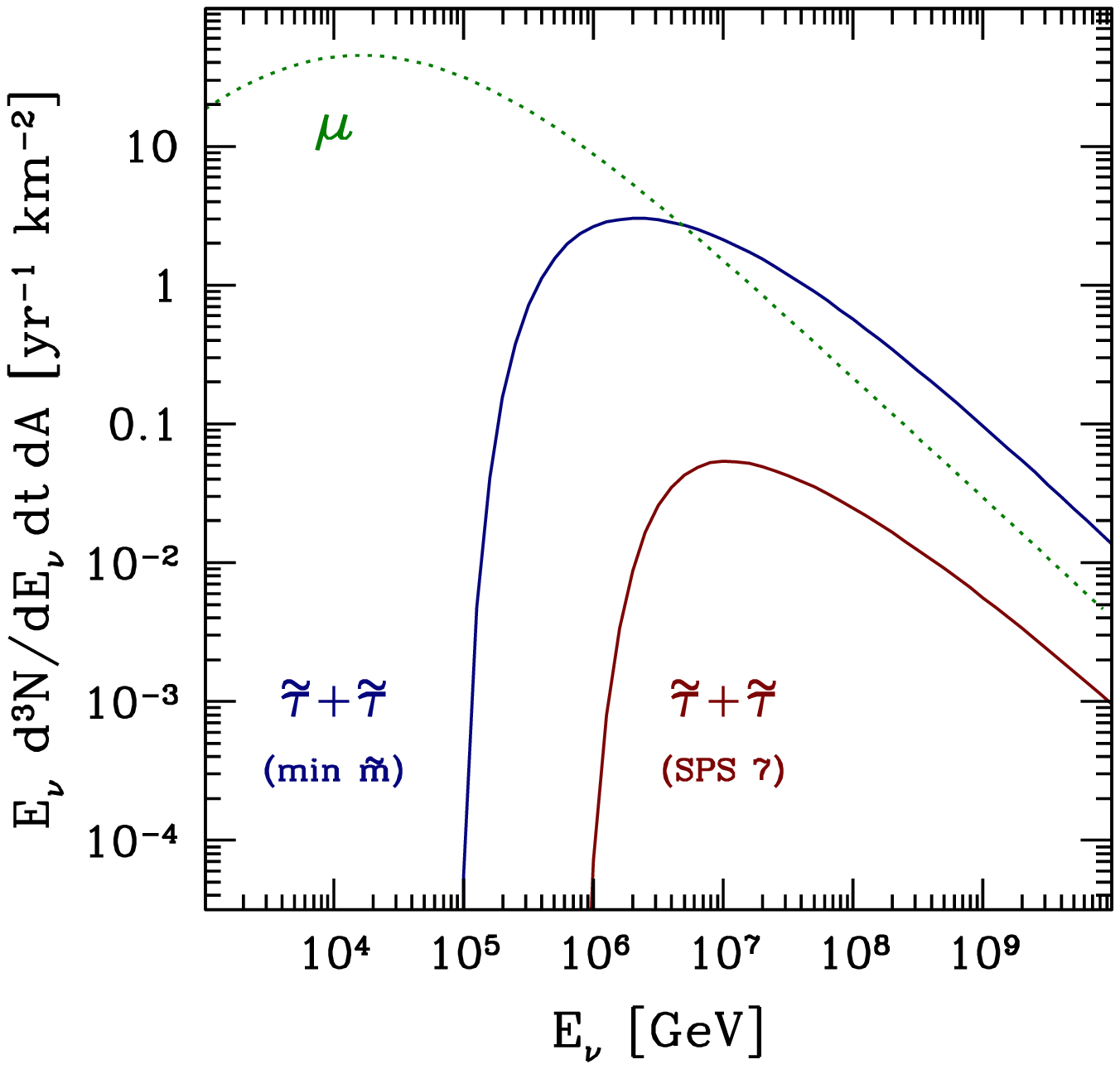}
\caption[]{\label{flux}The rate of parallel stau tracks through a neutrino telescope like IceCube in terms of the neutrino energy.}
\end{minipage}
\end{center}
\end{figure}

\section{Production and Propagation of Stau Pairs}

Cosmic neutrinos are exposed to SM charged and neutral current interactions in the Earth matter before they might undergo a supersymmetric process at some distance $l$ to the observer. We take these interaction into account by an exponential attenuation of the initial diffuse neutrino flux $F_\nu$ along the line of sight $-dz = n(\vec{x}) dl$ using the Earth's nucleon density $n=\rho/m_\text{p}$ shown in Ref.~\cite{Gandhi:1995tf}. The integrated column depth of the Earth, $z_\oplus$, depends on the nadir angle $\theta$ and the depth of the detector center below the horizon, which we fix at $1.9$~km corresponding to the IceCube detector~\cite{Ahrens:2002dv}. For the (yet
unknown) high energy cosmic neutrino flux we will adopt as a benchmark the Waxman-Bahcall flux of neutrinos from optically thin sources given in Ref.~\cite{Waxman:1998yy}: $E_\nu^2\,{F_\text{\tiny WB}(E_\nu)} \approx 6\!\times\! 10^{-8}\,\text{cm}^{-2}\,\text{s}^{-1}\,\text{sr}^{\,-1}\,\text{GeV}$ (all flavor).

The leading order supersymmetric contribution in deep inelastic scattering (DIS) off nucleons consists of slepton $\widetilde{l}$ and squark $\widetilde{q}$ production via chargino $\chi$ or neutralino $\chi^0$ exchange between neutrinos and quarks. The partonic cross sections can be found in Refs.~\cite{Albuquerque:2003mi,Ahlers:2006pf}. We have analyzed two different SUSY mass spectra with a long-lived $\widetilde{\tau}_R$ NLSP which we denote by ``{min~$\widetilde{m}$}", a scenario with super-particle masses just above the experimental limits  $m_{\chi} = m_{\chi^0} = m_{\widetilde{l}} = 100$~GeV and $m_{\widetilde{q}} = 300$~GeV, and ``{SPS~7}" with SUSY masses corresponding to the benchmark point SPS~7~\cite{Allanach:2002nj}. The resulting neutrino-nucleon cross section is shown in Fig.~\ref{sigma}. Here and in the following we have used the CTEQ6D parton distribution functions $f_j(x,Q^2)$~\cite{Pumplin:2002vw}.

We assume that the emerging sleptons and squarks will promptly decay into the $\widetilde{\tau}_\text{R}$ NLSP with an average energy of  $\langle E_{\widetilde\tau}\rangle_{\widetilde{l}}/E_{\widetilde{l}} = {0.5}$ (${1.0}$) and $\langle E_{\widetilde\tau}\rangle_{\widetilde{q}}/E_{\widetilde{q}}= {0.3}$ (${0.5}$) for the ``{SPS~7}'' (``{min~$\widetilde{m}$}'') mass scenario, respectively. This determines the average energy loss range $\langle z_{\widetilde{\tau}}\rangle_{\widetilde{l}/\widetilde{q}}$ of the staus which has been studied in great detail in Refs.~\cite{Reno:2005si,Albuquerque:2006am,Huang:2006ie}. To first order, the stau range scales linearly with the mass $m_{\widetilde{\tau}}$ and is about three orders of magnitude larger than the range of muons with the same energy. In our calculations we have used an approximation of $z_{\widetilde{\tau}}$ from Ref.~\cite{Reno:2005si} for staus with energy $E_{\widetilde{\tau}}>4\!\times\!10^5\,\text{GeV}\!\times\!(m_{\widetilde{\tau}}/150\,\text{GeV})$. Below this energy ionization losses dominate and were used in the calculation. For a given nadir angle the maximal range $z_\text{max}$ of an observable stau pair is then the lesser of $z_\oplus$, $\langle z_{\widetilde{\tau}}\rangle_{\widetilde{l}}$, and $\langle z_{\widetilde{\tau}}\rangle_{\widetilde{q}}$.

The flux of stau pairs through the detector from the decay of the emerging slepton and squark is then given by
\begin{equation}\label{fstaus}
\frac{\D^{4} N_{\widetilde{\tau}+\widetilde{\tau}}}{\D t\,\D A\,\D\Omega\,\D E_\nu}\,\, \approx 
\sum_\text{parton j}\,\,\int\limits_{E_\text{min}}^{E_\text{max}} 
\!\!\D E_{\,\widetilde{l}}\int\limits_0^{z_\text{max}}
\!\!\D z \int\limits_{x_\text{min}}^1 
\!\!\D x\,\,{f_j(x,Q^2)}\!\times\!\frac{\D^{2} \sigma^\text{\tiny SUSY}}{\D x\, \D E_{\,\widetilde{l}}}\!\times\! F_\nu(E_\nu)\!\times\!\text{e}^{-(z_\oplus-z)\sigma^\text{\tiny SM}}.
\end{equation}
Fig.~\ref{flux} shows the result of Eq.~(\ref{fstaus}) for the two SUSY mass scenarios ``min~$\widetilde{m}$'' and ``SPS~7'' compared to the rate of single muon events through the detector.

\begin{figure}[t!]
\begin{center}
\begin{minipage}[t]{0.47\linewidth}
\includegraphics[width=\linewidth,clip=true]{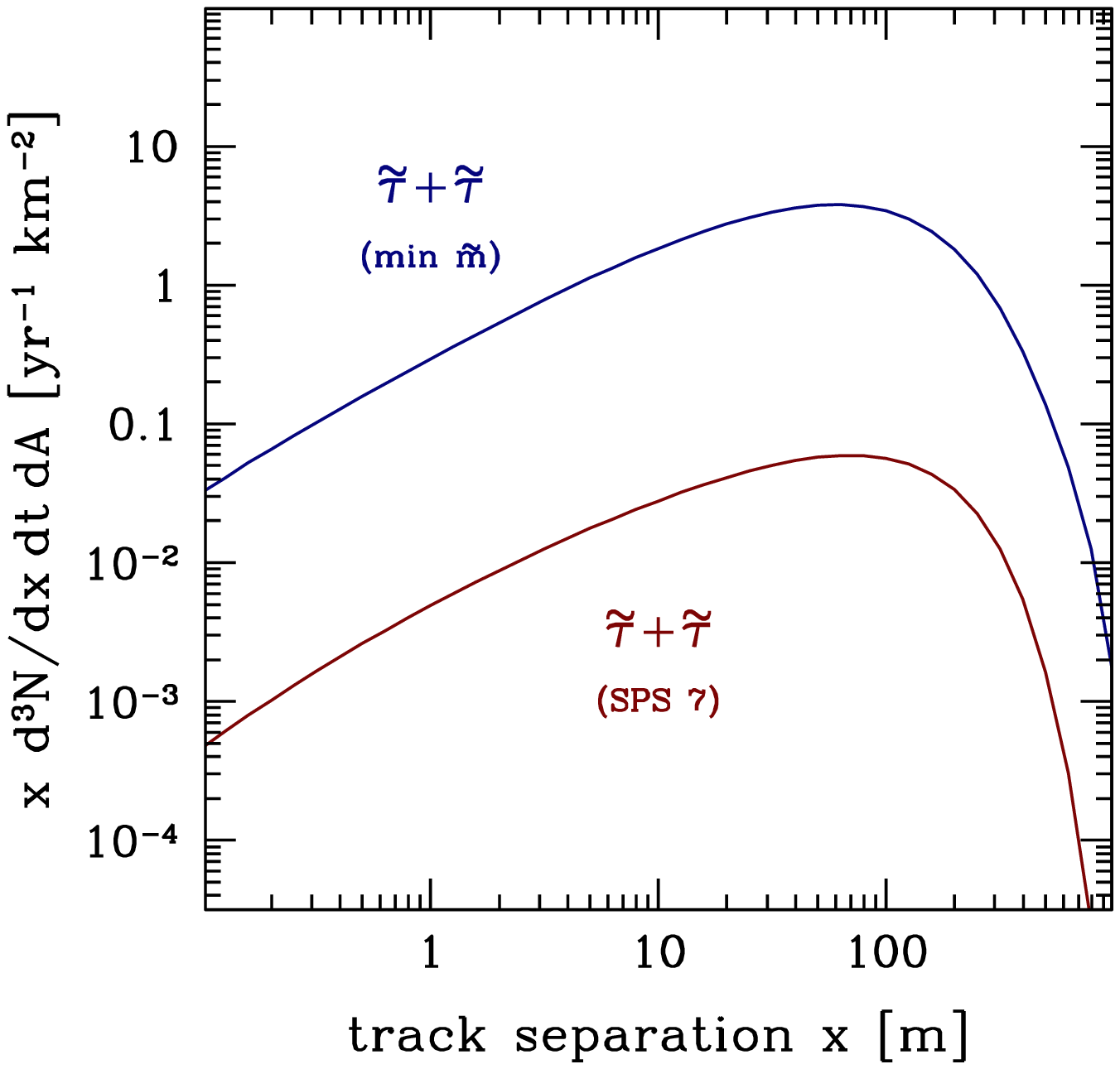}
\caption[]{\label{sep}The distribution of stau pairs with respect to their track separation in a Cherenkov detector.}
\end{minipage}
\hfill
\begin{minipage}[t]{0.47\linewidth}
\includegraphics[width=\linewidth,clip=true]{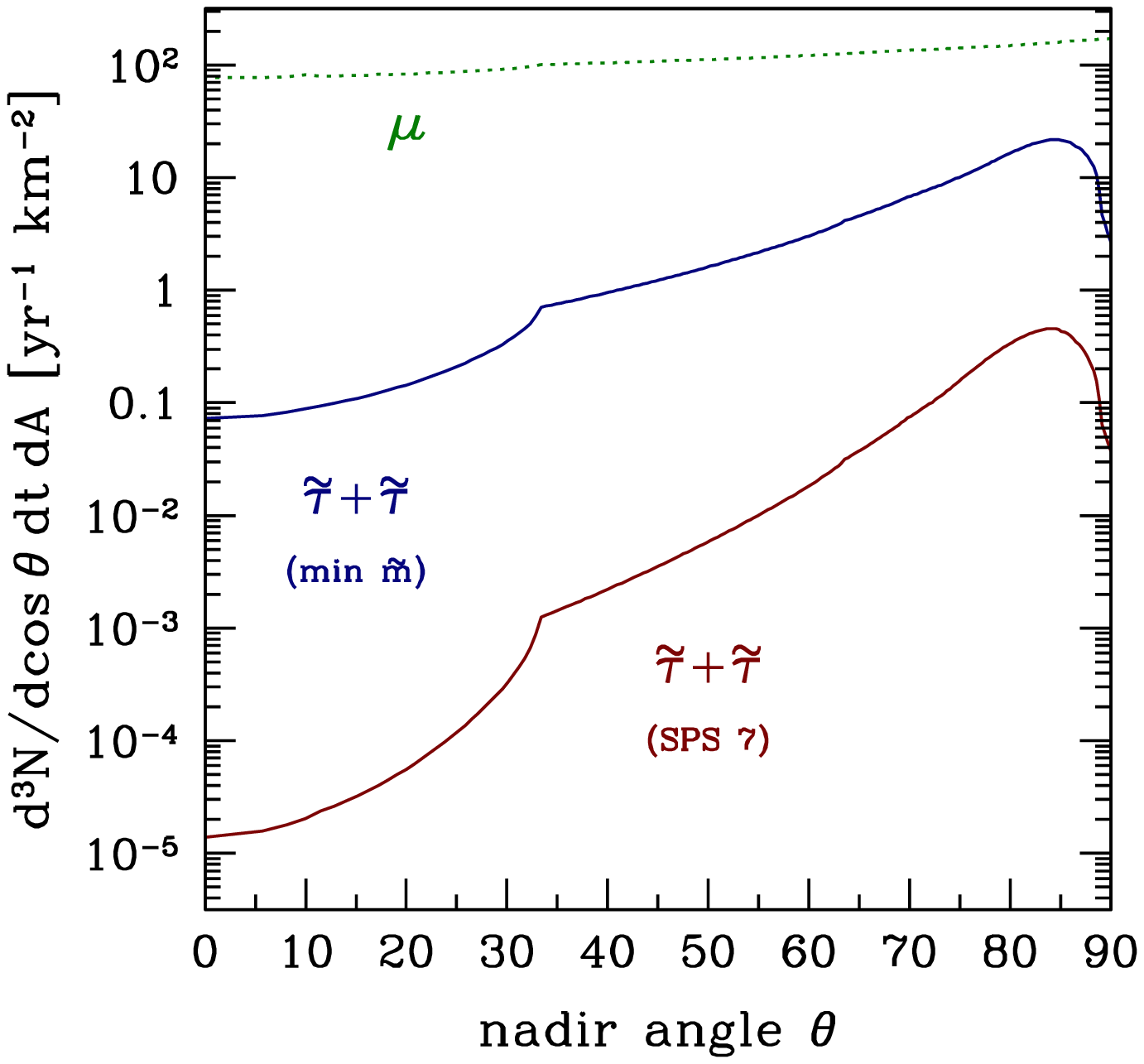}
\caption[]{\label{cos}The rate of stau pairs with respect to the nadir angle $\theta$ compared to single muon events.}
\end{minipage}
\end{center}
\end{figure}

\section{Detection Rate at Neutrino Telescopes}

Not all of the stau pairs (Eq.~(\ref{fstaus})) might be seen as separable tracks in a Cherenkov telescope if they emerge from interactions too close to the detector. As an estimate we use the opening angle $\delta$ between the slepton and squark in DIS to calculate the separation of stau tracks in the detector: $x=2\,l\tan(\delta/2)$. In Fig.~\ref{sep} we show the distribution of stau tracks according to their distance in the detector. In Tab.~\ref{rates} we show the absolute rate at a cubic kilometer neutrino telescope such as IceCube for different detector resolutions.

The distribution of stau pairs according to the nadir angle $\theta$ is shown in Fig.~\ref{cos}. It reflects the exponential attenuation of the flux of neutrinos depending on the integrated column depth of the Earth ($z_\oplus$) in the line of sight. The steepening of the differential rate below $33^\circ$ is an effect of the dense inner core of the Earth. Most of the stau pairs arrive from about $85^\circ$ with respect to nadir. The suppression at larger nadir angles is due to the small column depth of the Earth, $z_\oplus$, compared to the energy loss range of the staus. 

On completion the IceCube detector will consist of 80 strings separated by 125~m~\cite{Ahrens:2002dv}, where each string carries 60 optical modules separated by about 17~m. Since most of the stau pairs will arrive from about $5^\circ$ below the horizon an effective detector resolution of $50-100$~m with respect to their separation seems possible.

\begin{table}[t!]
\caption[]{\label{rates}Rates of upward-going stau pairs for two different SUSY mass scenarios and different detection thresholds of the track separation $x$.}
\begin{center}
\begin{tabular}{lcccc}
\br
\multicolumn{5}{c}{Rate of $\widetilde{\tau}$ pairs  $[\,\text{yr}^{-1} \,\text{km}^{-2}\,]$}\\
\mr
\makebox[1.7cm][l]{Scenario}&\makebox[1.7cm][c]{total}&\makebox[1.7cm][c]{$x>10$~m}&\makebox[1.7cm][c]{$x>50$~m}&\makebox[1.7cm][c]{$x>100$~m}\\
\mr
min~$\widetilde{m}$&12&9&5&2\\
SPS 7&0.19&0.14&0.07&0.03\\
\br
\end{tabular}
\end{center}
\end{table}

\section{Conclusions}

Long-lived charged particles from a new physics sector may produce alternative signals at Cherenkov telescopes even if their production cross section and branching ratio is very small. Attractive candidate scenarios include supersymmetry with $\widetilde\tau_{R}$ NLSPs and gravitino LSPs. For our choice of supersymmetric mass spectra and neutrino fluxes an observation of stau pairs at a rate of a few per year is possible (see Tab.~\ref{rates}). However, current bounds~\cite{Ribordy:2005fi} on the diffuse flux of high energy neutrinos are about one order of magnitude above the Waxman-Bahcall flux~\cite{Waxman:1998yy} and an improvement of the results by a factor $\sim$10 is possible.

\section*{Acknowledgements}
The author would like to thank I.~Albuquerque and Z.~Chacko for valuable discussions and J.~Kersten and A.~Ringwald for collaboration.


\section*{References}
\frenchspacing
\bibliography{refs}
\bibliographystyle{h-physrev3}

\end{document}